\documentclass[12pt]{article}
\usepackage{graphicx}
\usepackage{cite}
\usepackage{color}
\begin{document}
\title{How we discovered\\ the nonet of light scalar mesons}
\author{
Eef van Beveren\\
{\normalsize\it Centro de F\'{\i}sica Te\'{o}rica,
Departamento de F\'{\i}sica, Universidade de Coimbra}\\
{\normalsize\it P-3004-516 Coimbra, Portugal}\\
{\small http://cft.fis.uc.pt/eef}\\ [.3cm]
\and
George Rupp\\
{\normalsize\it Centro de F\'{\i}sica das Interac\c{c}\~{o}es Fundamentais,
Instituto Superior T\'{e}cnico}\\
{\normalsize\it Universidade T\'{e}cnica de Lisboa, Edif\'{\i}cio Ci\^{e}ncia,
P-1049-001 Lisboa, Portugal}\\
{\small george@ist.utl.pt}
}

\maketitle

\begin{abstract}
As has been confirmed meanwhile by lattice-QCD calculations
(see e.g.\ Ref. \cite{HEPLAT0702023}),
the confinement spectrum of non-exotic quark-antiquark
systems has its ground state for scalar mesons well above 1 GeV
in the Resonance Spectrum Expansion (RSE)\footnote{The RSE was designed for
the description of the complete resonance structure in meson-meson
scattering, for both the heavy- and light-quark sectors.}.
For instance, in the $S$-wave $K\pi$ RSE amplitude, a broad resonance
was predicted slightly above 1.4 GeV \cite{EPJC22p493},
which is confirmed by experiment as the $K^{\ast}_{0}$(1430).
However, a complete nonet of light scalar mesons was predicted
\cite{ZPC30p615}
as well, when a model strongly related to the RSE and initially
developed to describe the $c\bar{c}$ and $b\bar{b}$ resonance spectra
\cite{PRD21p772}
was applied in the light-quark sector. Thus, it was found that the
light scalar-meson nonet constitutes part of the ordinary meson spectrum,
albeit represented by {\it ``extraordinary''}
\cite{HEPPH0701038} poles \cite{EPJC22p493}.
Similar resonances and bound states appear in the charmed sector
\cite{PRL91p012003},
and are predicted in the $B$-meson spectrum
\cite{HEPPH0312078,MPLA19p1949}.

A recent work \cite{JPG34p1789}
confirmed the presence of light scalar-meson poles
in the RSE amplitude for
$S$-wave and $P$-wave $\pi\pi$ and $K\pi$ contributions
to three-body decay processes
measured by the BES, E791 and FOCUS collaborations.
\end{abstract}

\section{Scattering poles}

It is generally accepted that resonances in scattering
are represented by poles in the ``second'' Riemann sheet
of the complex energy plane \cite{2ndSheetPoles}.
Let us assume here that in a process of elastic
and non-exotic meson-meson scattering
one obtains scattering poles at
\begin{equation}
E\; =\; P_{0}\, ,\;\; P_{1}\, ,\;\; P_{2}\, ,\;\;\dots
\;\;\; .
\label{polepositions}
\end{equation}
Simple poles in $S$ may be considered
simple zeros in its denominator.
Hence, assuming a polynomial expansion,
we may \cite{AIPCP717p322,PoSHEP2005p108}
represent the denominator $D$ of $S$ by
\begin{equation}
D(E)
\;\propto\;
\left( E-P_{0}\right)\left( E-P_{1}\right)\left( E-P_{2}\right)\dots
\;\;\; .
\label{Sdenominator}
\end{equation}
Unitarity then requires that the $S$-matrix be given by\footnote{
Note that we do not consider here
a possible overall phase factor
representing a background
and stemming from the proportionality constant
in formula (\ref{Sdenominator}).}
\begin{equation}
S(E)
\; =\;
\frac{\textstyle\raisebox{5pt}{
$\left( E-P_{0}^{\ast}\right)\left( E-P_{1}^{\ast}\right)
\left( E-P_{2}^{\ast}\right)\dots$}}
{\textstyle\raisebox{-5pt}{
$\left( E-P_{0}\right)\left( E-P_{1}\right)\left( E-P_{2}\right)\dots$}}
\;\;\; .
\label{unitairS}
\end{equation}

If we assume that the resonances (\ref{polepositions})
stem from an underlying confinement spectrum,
given by the real quantities
\begin{equation}
E\; =\; E_{0}\, ,\;\; E_{1}\, ,\;\; E_{2}\, ,\;\;\dots
\;\;\; ,
\label{confinementpositions}
\end{equation}
then we may represent the differences $\left( P_{n}-E_{n}\right)$,
for $n=0$, 1, 2, $\dots$, by $\Delta E_{n}$.
Thus, we obtain for the unitary $S$-matrix the expression
\begin{equation}
S(E)
\; =\;
\frac{\textstyle\raisebox{5pt}
{$
\left( E-E_{0}-{\Delta E_{0}}^{\ast}\right)
\left( E-E_{1}-{\Delta E_{1}}^{\ast}\right)
\left( E-E_{2}-{\Delta E_{2}}^{\ast}\right)
\dots$}}
{\textstyle\raisebox{-5pt}
{$
\left( E-E_{0}-\Delta E_{0}\right)
\left( E-E_{1}-\Delta E_{1}\right)
\left( E-E_{2}-\Delta E_{2}\right)
\dots$}}
\;\;\; .
\label{unitairSdeltaE}
\end{equation}

\begin{figure}[htbp]
\begin{center}
\resizebox{!}{100pt}{\includegraphics{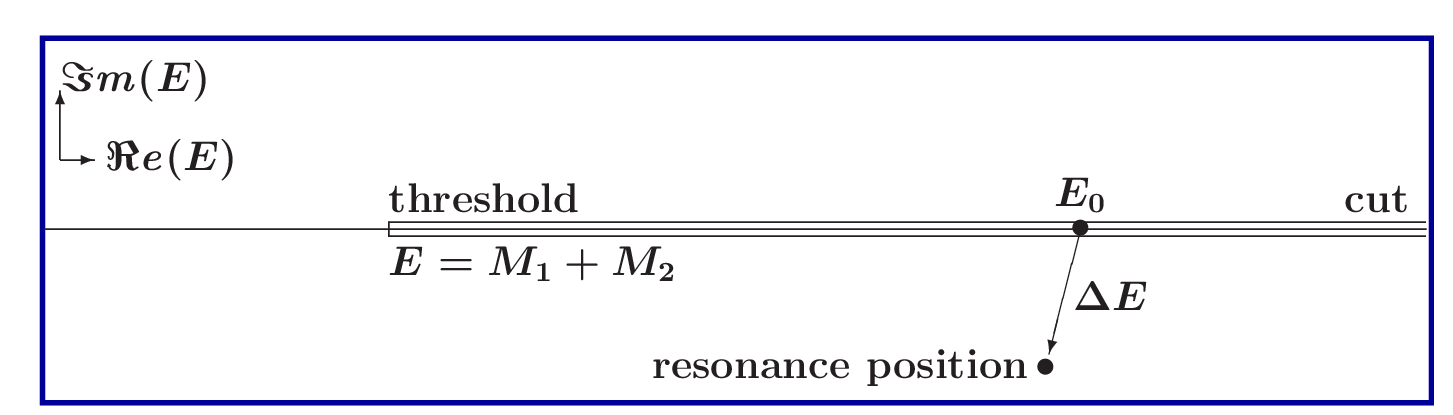}}
\end{center}
\vspace{-20pt}
\caption[]{\small
Illustration of a scattering pole
and the related energy level $E_{0}$ of the confinement
spectrum~(\ref{confinementpositions}).}
\label{poleposition}
\end{figure}

So we assume here that resonances occur in scattering because
the two-meson system couples to confined states, usually
of the $q\bar{q}$ type, viz.\ in non-exotic meson-meson scattering.
Let the strength of the coupling be given by $\lambda$.
For vanishing $\lambda$, we presume that the widths and real shifts
of the resonances also vanish (see Fig.~\ref{poleposition}).
Consequently, the scattering poles
end up at the positions of the confinement
spectrum~(\ref{confinementpositions}), and so
\begin{equation}
\Delta E_{n}\;\begin{array}{c} \\ \longrightarrow\\
\lambda\!\!\downarrow\! 0\end{array}\; 0
\;\;\;\;{\textstyle \mbox{\rm for}}\;\;\;\;
n\; =\; 0,\;\; 1,\;\; 2,\;\;\dots
\;\;\; .
\label{smalllambda}
\end{equation}
As a result, the scattering matrix tends to unity,
as expected in case there is no interaction.

An obvious candidate for an expression
of the form (\ref{unitairSdeltaE})
looks like
\begin{equation}
S(E)
\; =\;
\frac{\textstyle\raisebox{10pt}
{$
1+\lambda^{2}\,\left\{{\displaystyle \sum_{n}^{}}
\frac{\textstyle G(E)^{\ast}}{\textstyle E-E_{n}}\right\}
$}}
{\textstyle\raisebox{-10pt}
{$
1+\lambda^{2}\,\left\{{\displaystyle \sum_{n}^{}}
\frac{\textstyle G(E)}{\textstyle E-E_{n}}\right\}
$}}
\;\;\; ,
\label{unitairSRSE}
\end{equation}
where $G$ is a smooth complex function of energy $E$, and where,
at least for small values of the coupling constant $\lambda$,
one has
\begin{equation}
\Delta E_{i}\;\approx\;
-\lambda^{2}\,G\left( E_{i}\right)
\;\;\;\;{\textstyle \mbox{\rm for}}\;\;\;\;
i\; =\; 0,\;\; 1,\;\; 2,\;\;\dots
\;\;\; .
\label{deltaERSE}
\end{equation}
Relation (\ref{deltaERSE}) can be easily understood,
if we assume that for small $\lambda$
poles show up in the vicinity of
the energy values (\ref{confinementpositions})
of the confinement spectrum.
As a consequence,
at the zero $P_{i}$ of the denominator,
near the $i$-th recurrency of the confinement spectrum $E_{i}$,
the term $n=i$ dominates
the summations in formula (\ref{unitairSRSE}),
i.e.,
\begin{equation}
0
\; =\;
1+\lambda^{2}\,\left\{{\displaystyle \sum_{n}^{}}
\frac{\textstyle G\left( P_{i}\right)}{\textstyle P_{i}-E_{n}}\right\}
\;\approx\;
1+\lambda^{2}\,\frac{\textstyle G\left( E_{i}\right)}{\textstyle \Delta E_{i}}
\;\;\; .
\label{2ndorderRSE}
\end{equation}

For larger values of $\lambda$, one cannot
perform the approximation $P_{i}\approx E_{i}$
in Eq.~(\ref{2ndorderRSE}).
In such cases, the left-hand part of Eq.~(\ref{2ndorderRSE})
must be solved by other methods, usually numerically.
However, since it is reasonable to assume that poles
move smoothly in the lower half of the complex energy plane
as $\lambda$ varies, we may suppose that the left-hand part of
Eq.~(\ref{2ndorderRSE}) has solutions which,
when the value of $\lambda^{2}$ is continuously decreased,
each correspond to one of the values out of
the confinement spectrum (\ref{confinementpositions}).

When all scattering poles in expression~(\ref{unitairSdeltaE})
are known, one can --- with unlimited accuracy --- determine
the function $G$ in formula~(\ref{unitairSRSE}).
Once $G$ is known, one can search for poles by solving
the left-hand part of Eq.~(\ref{2ndorderRSE}).
However, further restrictions can be imposed upon
expression~(\ref{unitairSRSE}).
For a two-meson system, there may exist bound states
below the meson-meson scattering threshold.
Such states are represented by poles
in the analytic continuation of expression~(\ref{unitairSRSE})
to below threshold, on the real axis in the complex energy plane.
Consequently, in the case that a confinement state, say $E_{0}$,
comes out below threshold, its corresponding pole
is, at least for small coupling, expected to be found
on the real axis in the complex energy plane.
Using formula~(\ref{deltaERSE}), we obtain
\begin{equation}
G\left( E_{0}\right)\;\;{\textstyle \mbox{\rm real}}
\;\;\;\;{\textstyle \mbox{\rm for}}\;\;\;\;
E_{0}\, <\,{\textstyle \mbox{\rm threshold.}}
\label{Gbelowthreshold}
\end{equation}

Moreover, in order to ensure that scattering poles come out in the
lower-half of the complex energy plane, also using
formula~(\ref{deltaERSE}), we find that above threshold $G$ must be
complex, with a positive imaginary part.

\section{Partial waves}

In different partial waves, resonances come out at different masses.
At threshold, where the total invariant mass of the two-meson system
equals the sum of the two meson masses, one has additional conditions.
For $S$ waves, since cross sections are finite, we must demand that
$G$ do not vanish at threshold, whereas, for $P$ and higher waves, as
cross sections do vanish, $G$ should vanish as well.

A possible expression that satisfies all imposed conditions reads
\begin{equation}
i\, pa\, j_{\ell}(pa)\, h^{(1)}_{\ell}(pa)
\;\;\;\;{\textstyle \mbox{\rm for}}\;\;\;\;
\ell\; =\; 0,\;\; 1,\;\; 2,\;\;\dots
\;\;\; ,
\label{ijh1}
\end{equation}
where $p$ represents the linear momentum in the two-meson system
and $a$ a scale parameter with the dimensions of a distance.
The well-known scattering solutions $j_{\ell}$ and $h^{(1)}_{\ell}$
stand for the spherical Bessel function and the Hankel function of the
first kind, respectively.

Thus, we arrive at a good candidate for a scattering amplitude of
resonant scattering off a confinement spectrum, reading
\begin{equation}
T_{\ell}(E)
\; =\;
\frac{1}{2i}\left( S_{\ell}(E)-1\right)
\; =\;
\frac{\textstyle\raisebox{12pt}
{$
-2\lambda^{2}\,\left\{{\displaystyle \sum_{n}^{}}
\frac{\textstyle\raisebox{5pt}{$g_{n\ell}^{2}$}}
{\textstyle\raisebox{-5pt}{$E-E_{n}$}}\right\}\,
\mu pa\, j^{2}_{\ell}(pa)
$}}
{\textstyle\raisebox{-12pt}
{$
1+2i\lambda^{2}\,\left\{{\displaystyle \sum_{n}^{}}
\frac{\textstyle\raisebox{5pt}{$g_{n\ell}^{2}$}}
{\textstyle\raisebox{-5pt}{$E-E_{n}$}}\right\}\,
\mu pa\, j_{\ell}(pa)\, h^{(1)}_{\ell}(pa)
$}}
\; ,
\label{unitairTellRSE}
\end{equation}
where we have introduced the two-meson reduced mass $\mu$
and, moreover, relative couplings $g_{n\ell}$, which may be
different for different recurrencies of the confinement spectrum.

As it is written, formula~(\ref{unitairTellRSE}) seems to allow a
lot of freedom, through adjustments of the $g_{n\ell}$ to experiment.
In principle, it might even be useful to carry out such data fitting,
so as to gain more insight into the details of the coupling between
a two-meson system and a confined $q\bar{q}$ state.
However, experimental results are so far much too incomplete to
make a detailed comparison to our expression possible.

The spin structure of quarks, besides being important for the spectrum
of a $q\bar{q}$ system, is also crucial for the short-distance dynamics,
hence for the properties of the coupling between $q\bar{q}$ and meson-meson
states. In the $^{3\!}P_0$ model \cite{PRD8p2223,AP124p61},
it is assumed that a two-meson system couples to a $q\bar{q}$ state
via the creation or annihilation of a new $q\bar{q}$ pair, with vacuum
quantum numbers $J^{PC}=0^{++}$.
Under this assumption, all relative couplings can be determined
from convolution integrals of the wave functions. In
Refs.~\cite{ZPC21p291,PLB454p165}, such integrals have been 
calculated for general quantum numbers, including flavour.
The latter results leave no freedom for the coupling constants in
formula~(\ref{unitairTellRSE}), except for an overall strength $\lambda$,
which parametrises the probability of $q\bar{q}$ creation/annihilation.

This way, the full spin structure of the two-meson system
is entirely contained in the relative coupling constants $g_{n\ell}$.
Yet, direct comparison of the results given in
Refs.~\cite{ZPC21p291,PLB454p165} to experiment
would still be of great interest.

The relevant
$q\bar{q}\leftrightarrow MM$ coupling-constant book-keeping
has been developed in Refs.~\cite{ZPC21p291,PLB454p165}.
The latter scheme not only eliminates any freedom,
but also --- by construction --- restricts the number of possible
$MM$ channels that couple to a given $q\bar{q}$ system.
Nonetheless, the number of involved channels rapidly grows for
higher radial and angular excitations of the $q\bar{q}$ system.

\section{ Observables}

The scattering matrix is not directly observable,
but only through quantities like cross sections
and production rates.
It is straightforward to determine cross sections
\cite{PRD27p1527}
and, after some algebra, production rates
\cite{ARXIV07064119}
from expression (\ref{unitairTellRSE}).
However, a complete modelling of strong interactions is more complex.
For example, a $c\bar{c}$ vector state couples, via OZI-allowed decay,
to $D\bar{D}$, but also to $DD^{\ast}$, $D^{\ast}D^{\ast}$,
$D_{s}\bar{D}_{s}$, \ldots\ \cite{PRD21p772}.
Consequently, the involved two-meson channels couple to one another as
well.  So the first extension necessary for a more proper description
of strong interactions is the formulation of a multichannel
equivalent of expression~(\ref{unitairTellRSE}).
This issue has been dealt with in Ref.~\cite{AIPCP687p86}.
It involves coupling constants similar to the ones discussed above,
but now for each two-meson channel.

A meson-meson channel is characterised by quantum numbers,
including flavour and isospin, and the meson masses.
However, many of the needed masses are unknown yet, while most
mesons only exist as resonances.

In experiment, one can concentrate on one specific channel.
On the other hand, in a meaningful analysis all channels that couple
must be taken into account.
For example, one may argue that for the description of $\pi\pi$
scattering below the $KK$ threshold the channels $KK$, $\eta\eta$, \ldots\
can be neglected. But then one ignores virtual two-meson channels,
which may have a noticeable influence below the $KK$ threshold.

Furthermore, $q\bar{q}$ states may couple to one another via meson
loops. Typical examples are: $c\bar{c}$ vector states, which become
mixtures of $^{3\!}S_1$ and $^{3\!}D_1$ via loops of charmed mesons, and
isoscalar $q\bar{q}$ states, where kaon loops mix the $u\bar{u}+d\bar{d}$
and $s\bar{s}$ components. One then obtains different interplaying
confinement spectra, which may become visible in production rates.
The extension of expression~(\ref{unitairSRSE}) to more than one
$q\bar{q}$ channel has been considered in Refs.~\cite{ZPC30p615,PLB641p265},
for the description of the $\sigma$ and $f_{0}$(980) resonances.

\section{The parameters}

Besides the parameters $\lambda$ and $a$, formula~(\ref{unitairTellRSE})
contains an infinite number of parameters $E_{n\ell}$.
These represent the unknown and even hypothetical spectra of
confined $q\bar{q}$ systems.
From experiment, we only have data at our disposal
for resonances in meson-meson scattering or production.
Formulae like expression~(\ref{unitairTellRSE})
are intended to interpolate between the observed resonances and
the underlying --- largely unknown --- confinement spectrum.

In Fig.~2 of Ref.~\cite{AIPCP814p143} (see Fig.~\ref{KpiScross}), we showed,
for $S$-wave isodoublet $K\pi$ scattering,
how cross sections determined by the use of formula~(\ref{unitairTellRSE})
vary with increasing values of the coupling $\lambda$.
For small $\lambda$, the nonstrange-strange ($n\bar{s}$) confinement
spectrum is well visible in the latter figure,
whereas for the model value of the $q\bar{q}\,\leftrightarrow$ meson-meson
coupling experiment is reproduced.

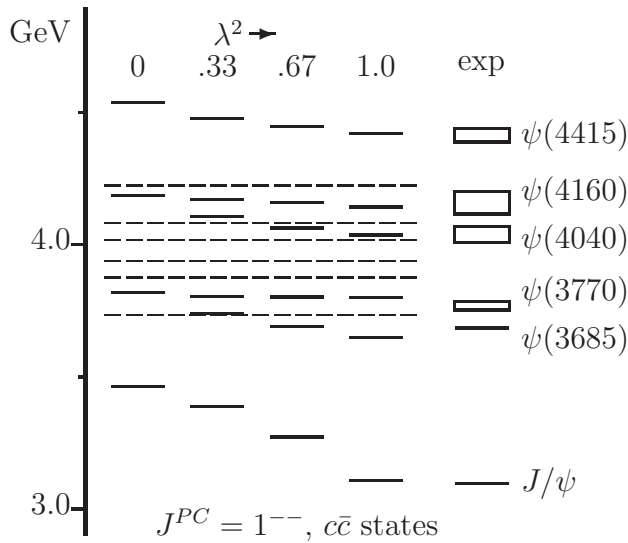
\begin{figure}[htbp]
\begin{center}
\begin{picture}(250,220)(-50,270)
\put(80,300){\makebox(0,0)[t]{$J^{PC}=1^{--}$, $c\bar{c}$ states}}
\linethickness{1.5pt}
\put(0,290){\line(0,1){200}}
\thicklines
\put(-5.0,300){\line(1,0){5}}
\put(-5.5,306){\makebox(0,0)[tr]{3.0}}
\put(-5.0,400){\line(1,0){5}}
\put(-5.5,406){\makebox(0,0)[tr]{4.0}}
\put(-2.5,350){\line(1,0){2.5}}
\put(-2.5,450){\line(1,0){2.5}}
\put(-5.5,485){\makebox(0,0)[tr]{GeV}}
\put(60,485){\makebox(0,0)[tr]{$\lambda^{2}$}}
\put(62,480){\vector(1,0){10}}
\put(20,464){\makebox(0,0)[b]{0}}
\put(50,464){\makebox(0,0)[b]{.33}}
\put(80,464){\makebox(0,0)[b]{.67}}
\put(110,464){\makebox(0,0)[b]{1.0}}
\put(150,464){\makebox(0,0)[b]{exp}}
\put(10,454.0){\line(1,0){20}}
\put(10,418.8){\line(1,0){20}}
\put(10,382.0){\line(1,0){20}}
\put(10,346.5){\line(1,0){20}}
\put(40,447.7){\line(1,0){20}}
\put(40,417.1){\line(1,0){20}}
\put(40,410.9){\line(1,0){20}}
\put(40,380.5){\line(1,0){20}}
\put(40,374.1){\line(1,0){20}}
\put(40,339.0){\line(1,0){20}}
\put(70,444.7){\line(1,0){20}}
\put(70,415.9){\line(1,0){20}}
\put(70,406.4){\line(1,0){20}}
\put(70,380.3){\line(1,0){20}}
\put(70,369.2){\line(1,0){20}}
\put(70,327.3){\line(1,0){20}}
\put(100,442.3){\line(1,0){20}}
\put(100,414.3){\line(1,0){20}}
\put(100,403.8){\line(1,0){20}}
\put(100,380.1){\line(1,0){20}}
\put(100,364.9){\line(1,0){20}}
\put(100,310.7){\line(1,0){20}}
\put(140,439.4){\framebox(20,4.3){}}
\put(165,441.5){\makebox(0,0)[lc]{$\psi$(4415)}}
\put(140,412.1){\framebox(20,7.8){}}
\put(165,419.9){\makebox(0,0)[lc]{$\psi$(4160)}}
\put(140,401.4){\framebox(20,5.2){}}
\put(165,401.9){\makebox(0,0)[lc]{$\psi$(4040)}}
\put(140,375.8){\framebox(20,2.4){}}
\put(165,382.0){\makebox(0,0)[lc]{$\psi$(3770)}}
\put(140,368.6){\line(1,0){20}}
\put(165,364.0){\makebox(0,0)[lc]{$\psi$(3685)}}
\put(140,309.7){\line(1,0){20}}
\put(165,309.7){\makebox(0,0)[lc]{$J/\psi$}}
\thinlines
\multiput(7.5,422.5)(8,0){15}{\line(1,0){6}}
\multiput(7.5,408.2)(8,0){15}{\line(1,0){6}}
\multiput(7.5,393.9)(8,0){15}{\line(1,0){6}}
\multiput(7.5,401.7)(8,0){15}{\line(1,0){6}}
\multiput(7.5,387.6)(8,0){15}{\line(1,0){6}}
\multiput(7.5,373.4)(8,0){15}{\line(1,0){6}}
\end{picture}
\end{center}
\vspace{-30pt}
\caption[]{\small The theoretical values of the central resonance positions
for charmonium S and D states for various values of the model parameter
$\lambda$, compared to the experimental situation. The various
dashed lines indicate the threshold positions of the strong decay
channels DD, DD$^{*}$, D$_{s}$D$_{s}$, D$^{*}$D$^{*}$,
D$_{s}$D$_{s}^{*}$ and D$_{s}^{*}$D$_{s}^{*}$.}
\label{charm}
\end{figure}
Furthermore, in Fig.~3 of Ref.~\cite{NPPS21p43} (see Fig.~\ref{charm})
we showed a similar
behaviour as a function of $\lambda$ for $J^{PC}=1^{--}$ $c\bar{c}$ states.
For $\lambda =0$, we find the theoretical ground state at 3.46 GeV,
whereas for $\lambda =1$ it coincides with the experimentally observed
$J/\psi$ mass.  The model employed to determine the results
of this figure was a multichannel extension of formula~(\ref{unitairTellRSE}),
taking moreover into account the degeneracy of certain confined $q\bar{q}$
states.

From these results we may conclude that, although there is some connection
between the confinement spectrum ($\lambda=0$)
and the resonances and bound states of two-meson systems ($\lambda=1$),
it is not a simple one-to-one relation.
Moreover, the level splittings of the confinement spectrum appear distorted
in experiment. In particular, the experimental ground states show up much
below the ground states of the hypothetical confinement spectrum.

Over the past decades, many models have been developed for the description
of meson spectra. Only very few of those models are based on
expressions for two-meson scattering or production. Here, it is stressed
that no data for the spectra of confined $q\bar{q}$ systems exist.
We only dispose of data for resonances in meson-meson
scattering or production \cite{PRD21p772,PRD27p1527,ZPC30p615}.
Nevertheless, in order to unravel the characteristics of the $q\bar{q}$
confinement spectrum, we must rely on results from experiment,
even though the available data \cite{JPG33p1} are manifestly insufficient
as hard evidence.

We observe from data that the average level splitting
in $c\bar{c}$ and $b\bar{b}$ systems equals 350--400 MeV,
when the ground states,
$J/\psi$, $\eta_{c}$, $\Upsilon (1S)$ and $\eta_{b}$
are not taken into account \cite{PRD21p772}.
Furthermore, mass differences in the
positive-parity $f_{2}$ meson spectrum,
which are shown in Table~3 of Ref.~\cite{EPJA31p468},
hint at level splittings of a similar size
in the light $q\bar{q}$ spectrum.
In Ref.~\cite{EPJA31p468}, possible internal flavour
and orbital quantum numbers for $f_{2}$ states were discussed.

Moreover, the few available mass differences for higher recurrencies
indicate that level splittings might turn out to be almost constant
for states higher up in the $q\bar{q}$ spectra as well
\cite{JPG33p1,ARXIV07073699}, a property shared by the spectrum
of a simple non-relativistic harmonic oscillator.
Over the past thirty years, we have systematically discussed
an ansatz for harmonic-oscillator confinement.
A formalism which naturally leads to a harmonic-oscillator-like
$q\bar{q}$ confinement spectrum starting from QCD,
by exploiting the latter theory's Weyl-conformal symmetry,
can be found in Refs.~\cite{NCA80p401,PRD30p1103}.

Guided by the --- not overwhelmingly compelling --- empirical evidence
that level splittings may be constant and independent of flavour,
and given the obvious need to further reduce the parameter  freedom in
expression~(\ref{unitairTellRSE}) for the two-meson elastic $T$-matrix,
we simply choose here the $q\bar{q}$ level splittings
$E_{(n+1)\ell}-E_{n\ell}$
to be constant and equal to 380 MeV,
for all possible $q\bar{q}$ flavour combinations.
The remaining set of parameters $E_{00}$,
different for each possible $q\bar{q}$ flavour combination,
can be further reduced \cite{PRD27p1527},
via the choice of effective valence flavour masses
and a universal frequency $\omega$.
In the future, when more data become available on the spectra of
$q\bar{q}$ systems, higher-order corrections to the harmonic-oscillator
spectrum may be inferred. At present, this does not seem to be feasible.
\clearpage

\section{\mbox{\boldmath $S$}-wave
scattering for \mbox{\boldmath $I\!=\!1/2$}}

\begin{figure}[htbp]
\begin{center}
\begin{tabular}{ccc}
\hline
\multicolumn{1}{|c|}{\includegraphics[height=150pt]{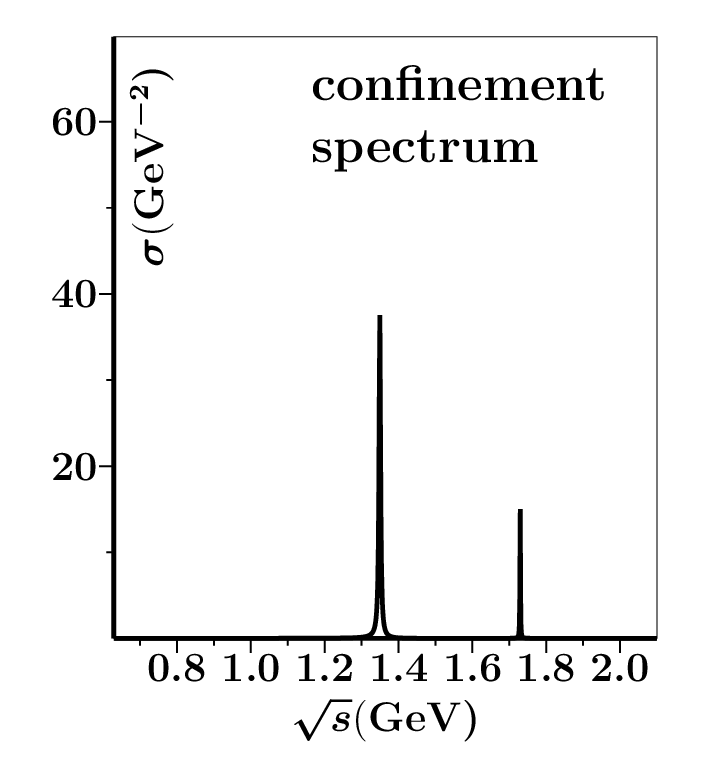}} &
\multicolumn{1}{|c|}{\includegraphics[height=150pt]{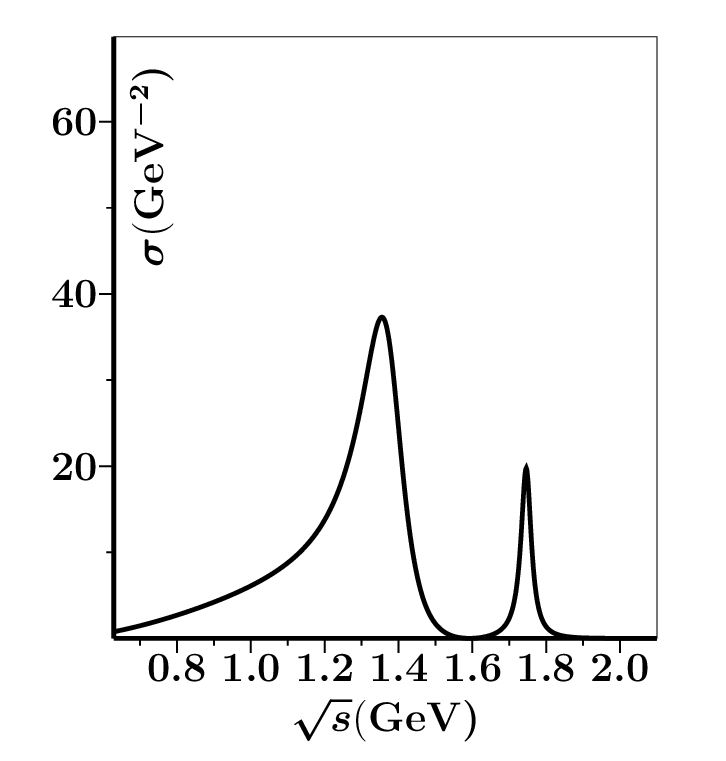}} &
\multicolumn{1}{|c|}{\includegraphics[height=150pt]{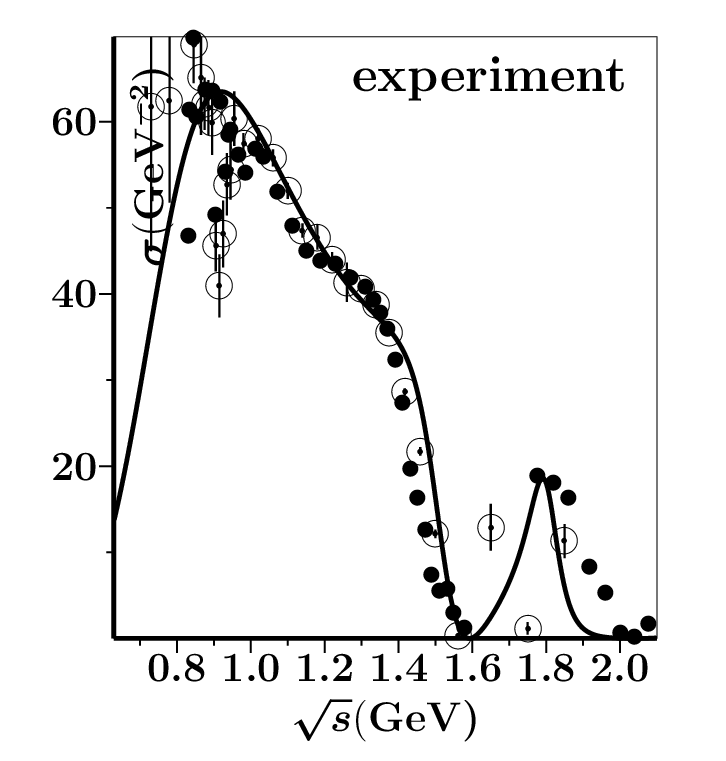}}\\
\hline
(a) & (b) & (c)\\
\end{tabular}
\end{center}
\vspace{-15pt}
\caption{\small Cross section for $S$-wave isodoublet $K\pi$ scattering.
Left: For very small values of $\lambda$, one observes
the $J^{PC}=0^{++}$ $n\bar{s}$ confinement spectrum.
Middle: When $\lambda$ takes about half its model value, one notices
some more structure for low invariant masses.
Right: At the model's value of $\lambda$, this structure is dominant and
well in agreement with the experimental observations.
The data are taken from Ref.~\cite{NPB133p490} (open circles)
and Ref.~\cite{NPB296p493} (full circles).
}
\label{KpiScross}
\end{figure}
In Fig.~2 of Ref.~\cite{AIPCP814p143} (see Fig.~\ref{KpiScross}),
we compared the result
of formula~(\ref{unitairTellRSE}) to the data of
Refs.~\cite{NPB133p490,NPB296p493}. We observed a fair agreement for
total invariant masses up to 1.6 GeV. However, one should bear in mind
that the LASS data must have larger error bars for energies above
1.5 GeV than suggested in Ref.~\cite{NPB296p493}, since most data points
fall well outside the Argand circle.
Hence, for higher energies, the model should better not follow the data
too precisely.

\begin{figure}[htbp]
\begin{center}
\begin{tabular}{ccc}
\hline
\multicolumn{1}{|c|}{\includegraphics[height=150pt]{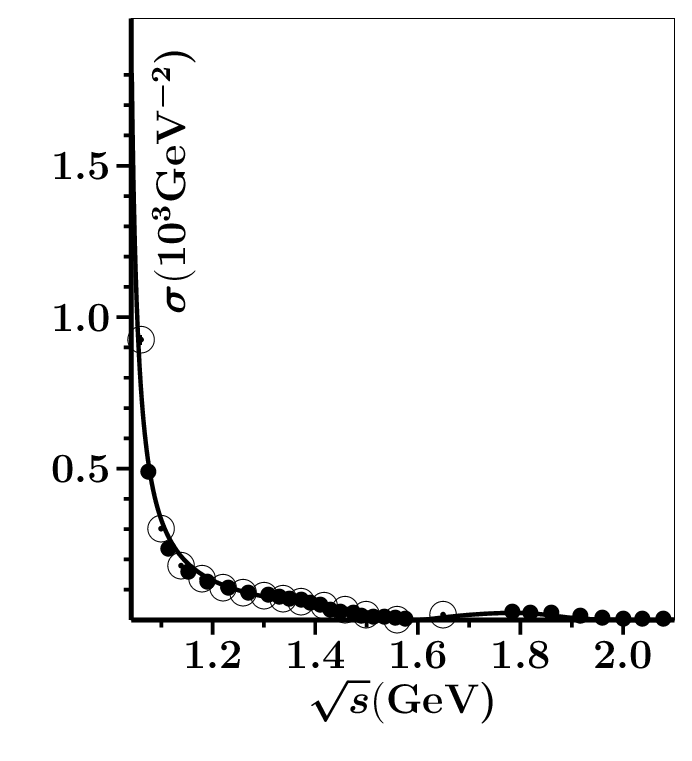}} &
\multicolumn{1}{|c|}{\includegraphics[height=150pt]{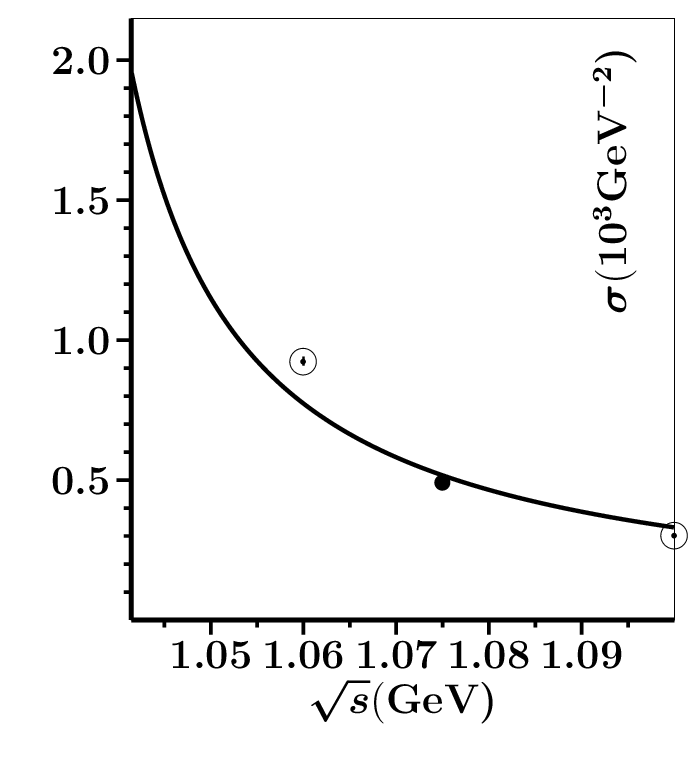}} &
\multicolumn{1}{|c|}{\includegraphics[height=150pt]{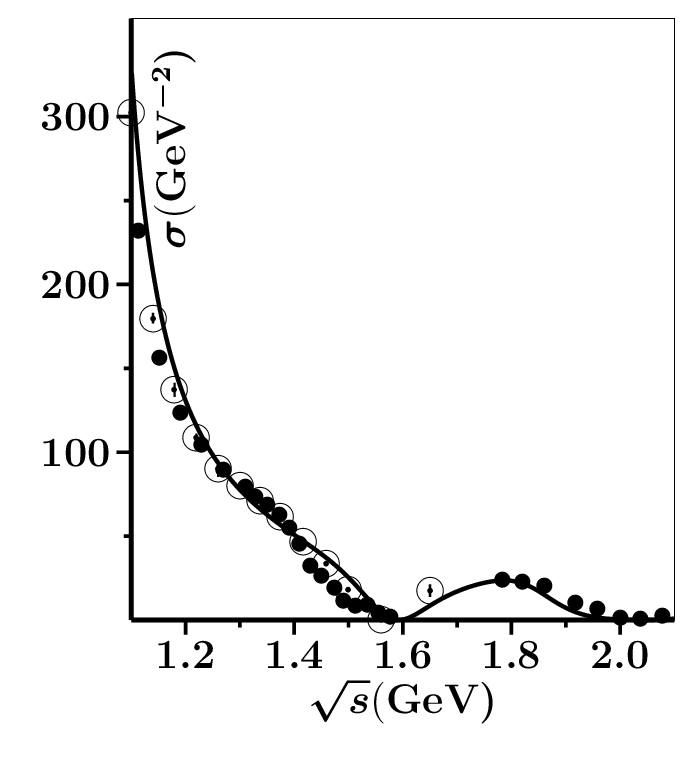}}\\
\hline
(a) & (b) & (c)\\
\end{tabular}
\end{center}
\vspace{-15pt}
\caption{\small $S$-wave $K\eta$ ``cross section'' (see text),
as a function of total invariant mass.
({\bf a}): From threshold up to 2.1 GeV.
({\bf b}): Detail for lower energy.
({\bf c}): Detail for higher energy.
}
\label{KetaS}
\end{figure}
\begin{figure}[htbp]
\begin{center}
\begin{tabular}{|c|}
\hline
\includegraphics[height=150pt]{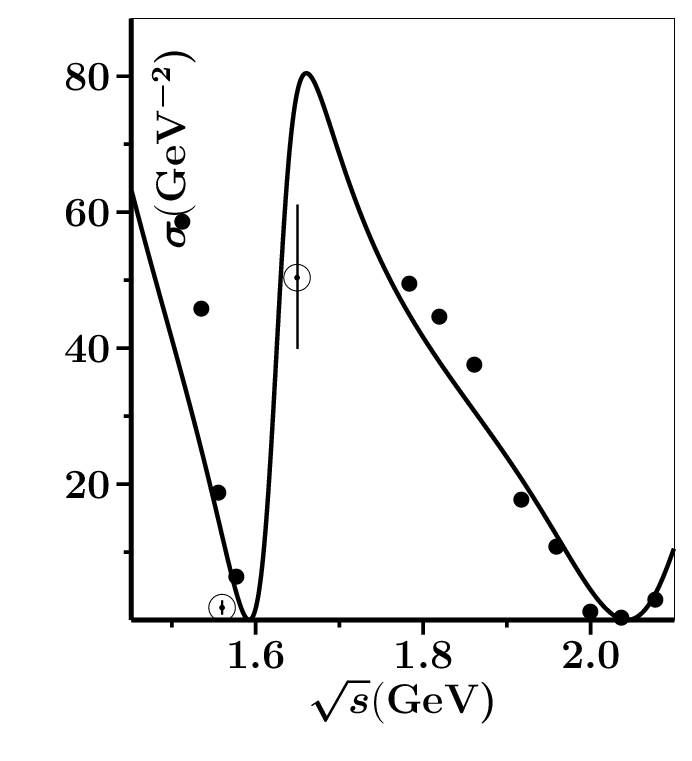}\\
\hline
\end{tabular}
\end{center}
\vspace{-15pt}
\caption{\small $S$-wave $K\eta'$ ``cross section'' (see text),
as a function of total invariant mass.
}
\label{KetapS}
\end{figure}

Now, in order to have some idea about the performance of
formula~(\ref{unitairTellRSE}) for $I\!=\!1/2\,$ $S$-wave $\pi K$ scattering,
we argue that, since in our model there is only one non-trivial eigenphase
shift for the coupled $\pi K$+$\eta K$+$\eta' K$ system, we may compare the
phase shifts of our model for $\eta K$ and $\eta' K$ to the experimental
phase shifts for $K\pi$. We did this comparison in Figs.~6 and 7 of
Ref.~\cite{AIPCP814p143}
(see respectively Figs.~\ref{KetaS} and \ref{KetapS}),
where, instead of the phase shifts, we plotted the
cross sections, assuming no inelasticity in either case. The latter assumption
is, of course, a long shot. Nevertheless, we observe an extremely good
agreement.

Apparently, we may conclude that the phase motion in the coupled $\pi K$,
$\eta K$ and $\eta 'K$ system is well reproduced by the model. In particular,
one could have anticipated that $\pi K$ and $\eta K$ have very similar phase
motions, because $\eta K$ has been observed to almost decouple from $\pi K$.
This implies that the corresponding $T$-matrix for a coupled
$\pi K$+$\eta K$ system is practically diagonal. Knowing, moreover, that this
system has only one non-trivial eigenphase, we should then also find almost
the same phase motion for $\pi K$ and $\eta K$.

\section{The \mbox{\boldmath $I\!=\!1/2\,$} \mbox{\boldmath $S$}-wave poles}

Since the model reproduces fairly well the data for the $I\!=\!1/2\,$
$S$-wave, it is justified to study its poles. In Table~\ref{Spoles} we collect
the five lowest zeros of formula (\ref{2ndorderRSE}).
\begin{table}[htbp]
\begin{center}
\begin{tabular}{||c|l||}
\hline\hline & \\ [-7pt]
Pole Position (GeV) & Origin \\
& \\ [-7pt]
\hline & \\ [-7pt]
$0.77-0.28i$ & continuum\\ [5pt]
$1.52-0.10i$ & confinement\\ [5pt]
$1.79-0.05i$ & confinement\\ [5pt]
$2.04-0.15i$ & continuum\\ [5pt]
$2.14-0.07i$ & confinement\\ [5pt]
\hline\hline
\end{tabular}
\end{center}
\caption[]{\small The five lowest zeros of formula (\ref{2ndorderRSE}).}
\label{Spoles}
\end{table}
Only three of the five corresponding poles are anticipated
from the $J^{P}=0^{+}$ $n\bar{s}$ confinement spectrum,
coming out at $1.39$ GeV, $1.77$ GeV, $2.15$ GeV, \ldots\ .
So we expected only three, but find five poles in the
invariant-mass region below 2.2 GeV. This shows that the transition
from formula~(\ref{unitairSdeltaE}) to formula~(\ref{unitairTellRSE}),
is not completely trivial. \em A forteriori, \em
expression~(\ref{2ndorderRSE}) even has more zeros than
expression~(\ref{Sdenominator}). It is amusing that Nature seems to agree
with the form of the scattering matrix in formula~(\ref{unitairTellRSE}).
As a matter of fact, the latter expression can be obtained by a model for
confinement \cite{PRD21p772,IJTPGTNO11p179}, whereas
formula~(\ref{unitairSdeltaE}) only expresses one of the many possible
ways to obtain poles in the scattering matrix at the
positions~(\ref{polepositions}).

The extra poles ({\it continuum} \/poles), which disappear towards negative
imaginary infinity when the overall coupling $\lambda$ is switched off,
can be observed in the experimental signal by noticing the shoulders at about
1.4 GeV in $\pi K$ scattering (see Fig.~2 of Ref.~\cite{AIPCP814p143}
(Fig.~\ref{KpiScross})), and
at about 1.9 GeV in $\eta 'K$ (see Figs.~6 and 7 of Ref.~\cite{AIPCP814p143}
(respectively Figs.~\ref{KetaS} and \ref{KetapS})).
The shoulder in $\pi K$ corresponds to the confinement state at 1.39 GeV,
on top of the larger and broader bump of the continuum pole at ($0.77-0.28i$)
GeV, while the shoulder in $\eta 'K$ corresponds to the continuum pole at
($2.04-0.15i$) GeV, on top of the larger and broader bump of the confinement
state at 1.77 GeV. Such subtleties in the data may have been overlooked
in the corresponding Breit-Wigner analyses.

There is one more observation to be made at this stage. The central
resonance peak of the lower enhancement in $S$-wave $\pi K$ scattering
(see Fig.~2 of Ref.~\cite{AIPCP814p143} (Fig.~\ref{KpiScross}))
is at about 830 MeV,
whereas the real part of the associated pole is at 772 MeV.
Hence, identifying the real part of the pole position with the central peak
of a resonance may be quite inaccurate.

With respect to the positions of the poles given in Table~\ref{Spoles},
it must be stressed again that these are model dependent.
So the model~(\ref{unitairTellRSE}) only indicates the
existence of such poles in the respective regions of total invariant
mass. A more sophisticated model, which fits the data even
better, will find the poles at somewhat different positions.

\section*{Acknowledgements}

This work was partly supported by the {\it Funda\c{c}\~{a}o para
a Ci\^{e}ncia e a Tecnologia} of the {\it Minist\'{e}rio da
Ci\^{e}ncia, Tecnologia e Ensino Superior} \/of Portugal,
under contract no.\ PDCT/FP/63907/2005.

\newcommand{\pubprt}[4]{#1 {\bf #2}, #3 (#4)}
\newcommand{\ertbid}[4]{[Erratum-ibid.~#1 {\bf #2}, #3 (#4)]}
\def\AIPCP{AIP Conf.\ Proc.}
\def\AP{Annals Phys.}
\def\EPJA{Eur.\ Phys.\ J.\ A}
\def\EPJC{Eur.\ Phys.\ J.\ C}
\def\IJTPGTNO{Int.\ J.\ Theor.\ Phys.\ Group Theor.\ Nonlin.\ Opt.}
\def\JPG{J.\ Phys.\ G}
\def\MPLA{Mod.\ Phys.\ Lett.\ A}
\def\NCA{Nuovo Cim.\ A}
\def\NPB{Nucl.\ Phys.\ B}
\def\NPPS{Nucl.\ Phys.\ Proc.\ Suppl.}
\def\PLB{Phys.\ Lett.\ B}
\def\PRD{Phys.\ Rev.\ D}
\def\PRL{Phys.\ Rev.\ Lett.}
\def\ZPC{Z.\ Phys.\ C}

\end{document}